\documentclass[prl,twocolumn,showpacs,preprintnumbers,amsmath,amssymb]{revtex4}

\usepackage{longtable}
\usepackage{graphicx}
\usepackage{dcolumn}
\usepackage{color}
\usepackage{bm}
\usepackage{mciteplus}
\graphicspath{{.},{./Ps/},{../Ps/},{../../Ps/}}
\begin{document}

\title{Shaping liquid drops by vibration}

\author{Andrey Pototsky} 
\affiliation{Department of Mathematics, Faculty of Science Engineering and Technology, Swinburne University of Technology, Hawthorn, Victoria, 3122, Australia}
\author{Michael Bestehorn}

\affiliation{Department of Theoretical Physics, Brandenburg University of Technology, 03044, Cottbus-Senftenberg, Germany}

\begin{abstract}
We present and analyze a minimal hydrodynamic model of a vertically vibrated liquid drop that undergoes dynamic shape transformations. In agreement with experiments, a circular lens-shaped drop is unstable above a critical vibration amplitude, spontaneously elongating in horizontal direction. Smaller drops elongate into localized states that oscillate with half of the vibration frequency. Larger drops evolve by transforming into a snake-like structure with gradually increasing length. The worm state is long-lasting with a potential to fragmentat into smaller drops.

\end{abstract}
\pacs{
68.15+e 
47.20.Ma 	
05.45.-a 	
}

\maketitle
Deformations of liquid drops, supported by a vertically vibrating solid plate, have been extensively studied in experiments over the period spanning almost seven decades. Responding to the vibration, a millimeter-sized circular drop oscillates and deforms, eventually taking a new shape. The variety of the final shapes, adopted by the drop is remarkably rich and the corresponding dynamical oscillation regimes are highly peculiar. At relatively small deformation amplitudes, one finds axisymmteric \cite{Holter52,*basaran_1992,*Yoshiyasu96} and non-axisymmetric \cite{MorihiroOkada2006,Noblin2009} polygonal vibrations, elongations, divisions, ejections of smaller drops \cite{MorihiroOkada2006} and bouncing or levitating drops \cite{Couder05}. 

A more dramatic change of the drop shape occurs as the result of the worm-like instability that leads to a significant elongation of the drop into a worm-like (or snake-like) twisting structure with a well defined width. Originally observed in drops floating on a more viscous immiscible fluid \cite{Pucci11,*Pucci13,*Pucci15}, the worm-like instability has recently been found in drops supported by a solid plate \cite{Hemmerle15}. 
Patterns of a similar shape resembling localized worms have been also found in layers of nematic liquid crystals near the onset of electroconvection driven by ac field \cite{Joets88,*Dennin96}.

Theoretical analysis of the shape of a three dimensional (3D) drop is only possible for quasi-stationary states with infinitesimally small oscillation amplitudes \cite{Rayleigh1879}. An alternative approach consists of assuming the balance between the capillary forces and radiation pressure of the surface waves to calculate the shape of an elongated steady drop \cite{Pucci11,*Pucci13,*Pucci15}. A phase field based model was earlier used to study swaying and spreading of liquid drops on vibrated plates in the regime of partial wetting \cite{Borc08,*Borc17}.

So far, the theory of the non-steady elongation dynamical regime in 3D was not available. To close this gap, we present in this letter a minimal model that accounts for the worm-like dynamical regimes of a vertically vibrated liquid drop.
\begin{figure}
\includegraphics[width=0.48\textwidth]{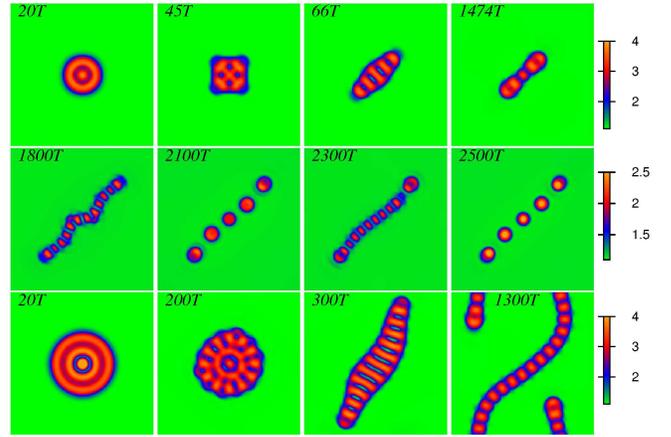}
\caption{\label{F1} (Upper row) development of a localized worm for $\langle h\rangle =1.19 L^2$ and $A=4$.  (Middle row) stretching and fragmentation at $A=9$. (Bottom row) Spatially periodic stable structure for $\langle h\rangle =1.41 L^2$ and $A=4$. Other parameters: $L=200$ $R=10$, $G=0.05$, $\Omega=0.05\pi$. }
\end{figure}
At finite Reynolds numbers, we employ the reduced model that was derived in the long-wave approximation from the Navier-Stokes equation to describe nonlinear Faraday waves in one-layer liquid films \cite{bes13,*bes13a}. The model consists of two coupled equations for the film thickness $h(x,y,t)$ and the flow ${\bm q}(x,y,t)=\int_0^h {\bm u}(z,x,y,t)\,dz$ across the film, where ${\bm u}$ is the horizontal fluid velocity. In the frame co-moving with the vibrated plate, the equations of motion are given 
\begin{eqnarray}
\label{eq1}
\frac{6\rho}{5}\left(\partial_t {\bm q} +{\bm \nabla}\cdot {\bm Q}-\frac{{\bm q}{\bm \nabla} \cdot {\bm q}}{7h}\right)&=&-\frac{3\eta {\bm q}} {h^{2}}-h{\bm \nabla} P\nonumber\\
\partial_t h+{\bm \nabla} \cdot {\bm q}&=&0,\nonumber\\
-\sigma \Delta h+\rho g(1+a(t))h&=&P,
\end{eqnarray}
with ${\bm Q}=9q_iq_k/7h$. In Eqs.\,(\ref{eq1}) $P$ is the pressure in the fluid, $\eta$, $\sigma$ and $\rho$ are the dynamic viscosity, surface tension and the density of the fluid, respectively. Time-periodic acceleration $g(1+a(t))=g(1+A\cos{\omega t})$ has frequency $\omega$ and a dimensionless amplitude $A$. 

A finite volume drop can be created by introducing a disjoining pressure that accounts for the partial wettability of the solid plate by the fluid. Thus, we extend the pressure term in Eqs.\,(\ref{eq1}) according to \cite{RevModPhys.57.827,kalliadasis2007thin} 
\begin{eqnarray}
\label{disj}
P=-\sigma \partial_{xx}h+\rho g(1+a(t))h+A h^{-3}\left(1-\left(h_0/h\right)^3\right),
\end{eqnarray}
with the Hamaker constant $A>0$ and the precursor film thickness $h_0$. Note that our model reduces to a vibrated Stokes drop \cite{uwe10} in the limit of a zero Reynolds number flow, i.e. when $q=h^3/3\eta\partial_x P$.

Next, we nondimenionalize Eqs.\,(\ref{eq1}) by scaling $h$ with $h_0$, time $t$ with $\eta \sigma h_0^5A^{-2}$, the coordinates $(x,y)$ with  $h_0^2\sqrt{\sigma A^{-1}}$ $h_0$ and the flow $q$ with $A^{3/2}/(\eta\sqrt{\sigma}h_0^2)$. The resulting dimensionless equations are obtained from Eqs.\,(\ref{eq1}) by setting $\eta=A=h_0=\sigma=1$ and replacing $6\rho/5$ with the Reynolds number $R=(6/5)\rho A^2/(\sigma \eta^2 h_0^3)$ and $\rho g $ with the gravity constant $G=\rho g h_0^4 A^{-1}$. The dimensionless vibration frequency is $\Omega = \eta \sigma h_0^5A^{-2} \omega$. The total volume of the liquid, including the volume of the  precursor film, is $V=\int_S h(x,y,t)\,dxdy$, where $S=L^2$ is the area of the plate, is conserved and can be used as a free parameter. For the remaining of the paper we fix $G=0.05$, $R=10$ and $\Omega=0.05\pi$ (vibration period of $T=40$).


In the absence of vibration, any stationary drop corresponds to zero flux $q(x,y)=0$. In this case, standard Stokes limit results apply \cite{kalliadasis2007thin}. An infinite-volume drop has a flat upper surface of the height $h_{\rm max}$. The drop rests on the precursor film of thickness $h_p$. Both $h_{\rm max}$ and $h_p$ are determined from the Maxwell equal-area construction to the effective pressure $\Pi(h)=Gh +h^{-3}-h^{-6}$ \cite{SM_max}. Thus, for $G=0.05$, we find $h_{\rm max}=4.03$ and $h_p\approx 1$.  A finite volume drop can be created if the average thickness of the liquid $\langle h \rangle = VS^{-1}$ is in the range $\langle h \rangle \in [h_p,1.95]$.  The drop co-exists with a linearly stable flat film with identical thickness $\langle h \rangle$ if $\langle h \rangle \in [h_p,1.3]$. Otherwise, the drop is absolutely stable, i.e. a flat film is linearly unstable due to dewetting.

In Fig.\ref{F1} we demonstrate three evolution scenario of a circular drop that undergoes a worm-like instability. A lens-shaped drop of relatively small volume \cite{volume} develops into a stable localized worm at $A=4$ (upper row). At time $t=1474T$ the vibration amplitude is abroptly increased to $A=9$. The worm begins to stretch and twist, eventually fragmenting into five isolated circular drops at $t=2100T$, as shown in Fig.\ref{F1}(middle row). The fragmented drops merge back into a compact worm at $t=2300T$ and then reappear later at $t=2500T$. Our simultions show that larger vibration amplitudes enhance chances of fragmentation. The final arrangement of the isolated drops after fragmentation may be rather complex, e.g. the drops may form pairs (dipols) \cite{SM_fragment}. Breaking of an elongated drop into several fragments was earlier observed in the experiments with floating drops \cite{Pucci11,*Pucci13,*Pucci15}.

The evolution of a drop with a relatively large volume is shown at $A=4$ in the bottom row in Fig.\ref{F1}. Initially the wobbling drop develops concentric circular surface waves that corresponds to the lowest energy axisymmetric mode \cite{Noblin2009,Hemmerle15}. As the instability grows, the vibration assumes an axisymmetric polygonal form that oscillates with half of the forcing frequency \cite{Holter52,*basaran_1992,*Yoshiyasu96}. Eventually, the drop elongates, developing a distinct worm-like structure. The length of the worm increases to reach the size of the domain. Further worm stretching is halted due to periodic boundary conditions. The newly formed S-shaped spatially periodic state is stable and does not fragment over the time window of $1000$ oscillation cycles.

We found that worm-like instability persists in drops with gradually increasing volume \cite{SM_dewetting}. A developing worm draws the liquid from the background film, whose thickness is larger than  $h_p\approx 1$.
This situation corresponds to drops, emerging from dewetting liquid layers.  

To estimate the characteristic length of the vibration patterns, we calculate the Faraday instability threshold of a an infinite-volume drop with flat upper surface of the height $h_{\rm max}=4.03$. 
By linearizing Eqs.\,(\ref{eq1}) around the flat film $h=h_{\rm max}+\delta h $ and zero flux $q=\delta q $ we obtain
\begin{eqnarray}
\label{eq3}
0&=&R \ddot{\delta h}+3 h_{\rm max}^{-2}\dot{\delta h}+h_{\rm max}B\delta h,
\end{eqnarray}
with $B=\left[k^2 + G (1+A\cos{(\Omega t)})-3h_{\rm max}^{-4}+6h_{\rm max}^{-7}\right]$.
For $\Omega=0.05\pi$ the Faraday waves on the surface of a flat film with thickness $h_{\rm max}=4.03$ set in at $A_c=1.38$ that corresponds to the tip of the first sub-harmonic tongue \cite{SM_max}. Note that the precursor film remains effectively stable (it becomes Faraday unstable for $A>80$).
The length of the least stable Faraday wave $\lambda=18$ (tip of the tongue) coincides well with the typical size of the ripples in Fig.\,\ref{F1} (bottom row).
 
\begin{figure}
\includegraphics[width=0.49\textwidth]{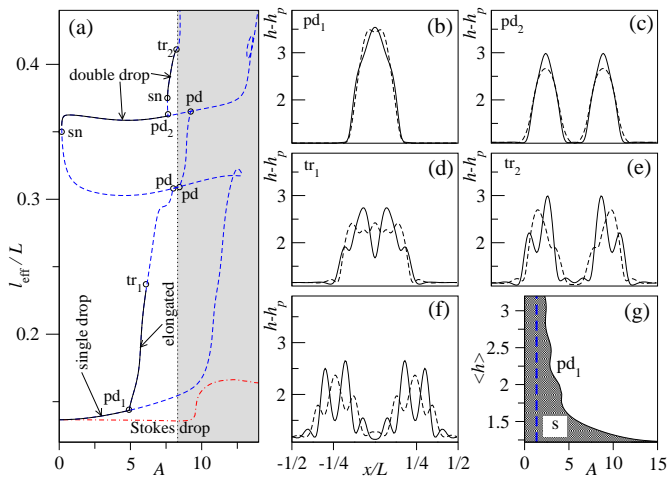}
\caption{\label{F2} (a) Bifurcation diagram of a wobbling drop with $\langle h \rangle=1.6$ and $L=100$. Other parameters as in Fig.\ref{F1}. Effective size of the drop $l_{\rm eff}/L$ as a function of the vibration amplitude $A$. Solid lines correspond to stable oscillatory states. Period doubling, torus and saddle-node bifurcation points are marked by {\tt pd}, {\tt tr} and {\tt sn}, respectively.  Dotted-dashed line corresponds to a linearly stable wobbling Stokes drop, i.e. $R=0$. In the shaded localized drops are not found. (b,c,d,e) Snapshots of stable localized solutions taken from points ${\rm pd}_1$, ${\rm pd}_2$, ${\rm tr}_1$, ${\rm tr}_2$ with oscillation periods: $T$ (b,d) and $2T$ (c,e).  (f) Spatially periodic solution at $A=8.5$ obtained from direct numerical solution of Eqs.\,(\ref{eq1}). (g) Locus of the ${\rm pd}_1$ points in space $(A,\langle h\rangle)$. An oscillating drop with period $T$ is stable in the shaded area. Dashed vertical line corresponds to the onset of the Faraday waves on surface of a flat film with height $h_{\rm max}$. 
}
\end{figure}
%
 Our next goal is to calculate the linear stability of a wobbling lens-shaped drop, detect possible bifurcations and describe how the shape of the drop changes, when the vibration amplitude is gradually increased. The stability of a two-dimensional (2D) drop $h(x,t)$ can be analyzed using numerical continuation technique \cite{AUTO,SM_cont}. We then proceed to show that the results in 2D are in qualitative agreement with the 3D drops.



We set $\langle h\rangle=1.6$ to create a lens-shaped drop in the domain $L=100$ and obtain the stationary drop $h_s(x)$ from the direct numerical simulation of Eqs.\,(\ref{eq1}) in 2D in the absence of vibration, i.e. $A=0$. Next, we numerically continue this solution using the forcing amplitude $A$ as the continuation parameter.  
The effective size $l_{\rm eff}$ of a drop that oscillates with a period $\tau$ is found according to
$l_{\rm eff}=\left\langle \sqrt{\overline{x^2}}\right\rangle_\tau$, 
%
with $\overline{ x^2}=(V-h_p L)^{-1}\int_{-L/2}^{L/2}dx\,x^2 (h(x,t)-h_p)$ and $\langle\dots\rangle_\tau=\tau^{-1}\int_{0}^{\tau}dt(\dots)$.

Similar to the Faraday waves on the surface of a flat film, there exist two kinds of oscillating drops: the harmonic drops that oscillate with the period of $\tau=T=2\pi/\Omega$, and the subharmonic drops that oscillate with the period of $\tau=2T$. 

We find that at small amplitudes $A$, the drop wobbles with the period of the external vibration $T$. The shape of the 2D drop remains symmetric at all times, as shown in Fig.\ref{F2}(b), where we plot two profiles $h(x,t)$ that correspond to a maximum and a minimum drop height. 

 A wobbling 2D drop becomes linearly unstable at $A_c=4.95$ with respect to a period-doubling bifurcation (point marked by ${\rm pd}_1$ in Fig.\ref{F2}(a)). The new branch, bifurcating at ${\rm pd}_1$, corresponds to subharmonic solutions that oscillate with twice the period of the forcing, i.e. $\tau=2T$. These new solutions correspond to a horizontally elongated drop, as shown in Fig.\ref{F2}(d).

At $A=6.08$ the elongated state becomes unstable as the result of the torus bifurcation (point marked by ${\rm tr}_1$) \cite{torus}. 
There exist two new kinds of stable solutions, associated with an oscillating double drop. The first kind of the double drop is harmonic, oscillating with the period $\tau=T$. Two snapshots of $h(x,t)$ from point ${\rm pd}_2$ with the minimal and the maximal heights are plotted in Fig.\ref{F2}(c).  The shape of each of the two drops in the double drop state remains symmetric at all times. The branch of the harmonic double drops stretches to vanishingly small vibration amplitudes, implying that such a double drop co-exists with the single drop and with the polygonal state.

The second kind of the double drop is sub-harmonic, oscillating with the period $\tau=2T$, as shown in Fig.\ref{F2}(e). The branch of these solutions bifurcates from the branch of the harmonic double drops as the result of the period-doubling bifurcations at point ${\rm pd}_2$. The shape of the two smaller drops in a sub-harmonic double drop changes in time in a non-symmetric way: the two drops oscillate in a cycle by moving towards and then away from each other. Such configuration is also found in 3D \cite{SM_fragment}.

The double- and a the single-drop states belong to one solution branch that snakes its way upwards without breaking in the diagram in Fig.\ref{F2}(a). This feature is reminiscent of the so-called snaking behavior, when steady localized structures are formed by spatially periodic patterns that are connected to a homogeneous solution \cite{CrHo93,sandstede10}. In system with larger liquid volume we also found multiple drop states, i.e. triple and quadruple drops, etc (not shown here). 

\begin{figure}
\includegraphics[width=0.49\textwidth]{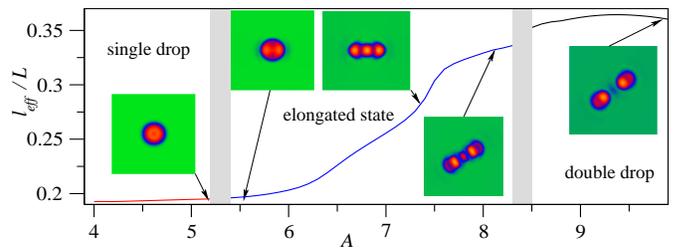}
\caption{\label{F3} 
Shape transformations of a 3D drop with $\langle h\rangle=1.2$ and $L=100$ during numerical continuation for gradually increasing vibration amplitude $A$, as described in the main text. Effective size in 3D $l_{\rm eff}=\left\langle \sqrt{\overline{x^2+y^2}} \right\rangle_\tau$. Transition from a circular to elongated shape at $A\approx5.1$ is a period-doubling bifurcation. 
}
\end{figure}

At the point marked by ${\rm tr}_2$ the sub harmonic double drop looses its stability via a torus bifurcation. Beyond the ${\rm tr}_2$ point, in the grey-shaded area Fig.\ref{F2}(a), the localized states in 2D no longer exist. At such large amplitudes, a single or a double drop is completely destroyed by the vibration.  In the direct simulations of Eqs.\,(\ref{eq1}) in 2D we either find spatially periodic oscillatory states as in Fig.\ref{F2}(f) obtained for $A=8.5$, or standing Faraday waves on the surface of a liquid film for $A>9.$ (not shown).


The role of the Reynolds number in the worm-like instability is clarified by calculating the stability of the wobbling Stokes drop (i.e. $R=0$), shown by the dotted-dashed line in Fig.\ref{F2}(a). We find that a Stokes drop remains linearly stable for any values of the vibration amplitude. 

In Fig.\ref{F2}(g) we plot the locus of the period-doubling bifurcation points ${\rm pd}_1$ {\it vs} $\langle h\rangle$. As the drop volume increases, the upper surface of the drop flattens and its height approaches $h_{\rm max}$. For large volume drops the elongation sets in at $A_c\approx 1.38$ (vertical dashed line) that corresponds to the tip of the first subharmonic tongue of the Faraday instability of a flat film with $h=h_{\rm max}=4.03$.  Presumably due to the stabilizing effect of the capillary forces at its edge \cite{douady_1990}, any finite volume drop is more stable than a flat film with identical height.  

Finally, we show that a 3D drop follows a similar bifurcation scenario as in Fig.\ref{F2}(a). We set $\langle h\rangle=1.2$ in the domain $L=100$ \cite{note_size} and solve Eqs.\,(\ref{eq1}) in 3D by employing a primitive continuation schedule: a single drop solution is followed by gradually increasing the amplitude from $A=4$ to $A=10$ with step $0.1$ and allowing each new solution to equilibrate over $300$ oscillation cycles. We find that a period $T$ wobbling drop develops an elongated state at $A\approx 5.1$, which then at $A\approx 8.1$ transforms into a double drop with oscillation period $2T$, as illustrated in Fig.\ref{F3}.

In conclusion, our model captures experimentally observed dynamic regimes \cite{Pucci11,*Pucci13,*Pucci15,Hemmerle15} of a vertically vibrated liquid drop with large deformation amplitudes.  Below the critical vibration amplitude, a circular lens-shaped drop wobbles with the frequency of the external vibration.  Horizontal elongation occurs as the result of a period-doubling bifurcation similar to the Faraday instability of a flat liquid surface. 
The final fate of the developing worm-like structure can be controlled by changing the drop volume, or by fine-tuning the vibration amplitude. In an unbounded domain, a localized worm can be stable, or may further fragment into smaller pieces. Self-interaction in case of periodic boundaries allows for stable spatially periodic worm-like structures.

\ifx\mcitethebibliography\mciteundefinedmacro
\PackageError{apsrevM.bst}{mciteplus.sty has not been loaded}
{This bibstyle requires the use of the mciteplus package.}\fi


\begin{mcitethebibliography}{31}
\bibitem[{vol()}]{volume}
\bibinfo{note}{The volume of a drop can be obtained from the total liquid
  volume $V=\langle h\rangle S$ by subtracting the volume $Sh_p$ of the
  precursor film.}\relax
\mciteBstWouldAddEndPunctfalse
\mciteSetBstMidEndSepPunct{\mcitedefaultmidpunct}
{}{\mcitedefaultseppunct}\relax
\EndOfBibitem
\bibitem[{tor()}]{torus}
\bibinfo{note}{New solution branch that bifurcates at point ${\rm tr}_1$ is
  unstable and is not shown here.}\relax
\mciteBstWouldAddEndPunctfalse
\mciteSetBstMidEndSepPunct{\mcitedefaultmidpunct}
{}{\mcitedefaultseppunct}\relax
\EndOfBibitem
\bibitem[{not()}]{note_size}
\bibinfo{note}{2D and 3D drops with identical average thickness $\langle
  h\rangle$ cannot be directly compared, since the horizontal size in 2D and 3D
  is different at fixed aspect ratio $L$.}\relax
\mciteBstWouldAddEndPunctfalse
\mciteSetBstMidEndSepPunct{\mcitedefaultmidpunct}
{}{\mcitedefaultseppunct}\relax
\EndOfBibitem
\end{mcitethebibliography}

\end{document}